\documentclass[usenatbib]{mn2e}
\usepackage{graphicx}
\usepackage{lscape}
\usepackage{epsfig}
\title{Galactic Satellite Systems: Radial
Distribution and Environment Dependence of Galaxy Morphology}
\author[H. B. Ann, Changbom Park, Yun-Young Choi]
{H. B. Ann$^{1}$\thanks{E-mail:hbann@pusan.ac.kr}, 
Changbom Park$^{2}$\thanks{E-mail:cbp@kias.re.kr}, 
Yun-Young Choi$^{3}$\thanks{
E-mail:yychoi@kias.re.kr}\\
$^{1}$ Division of Science Education, Pusan National
University, Busan 609-735, Korea,\\
$^{2}$ Korea Institute for Advanced Study, 
Dongdaemun-gu, Seoul 130-722, Korea\\
$^{3}$ Astrophysical Research Center for the Structure and 
Evolution of the Cosmos, Sejong University, 
Seoul 143-747, Korea}
\date{Accepted .      Received ;      in original form.. }
\pagerange{\pageref{firstpage}--\pageref{lastpage}}
\pubyear{2008}
  
\begin{document}

\maketitle

\label{firstpage}

\begin{abstract}
We have studied the radial distribution of the early (E/S0) and late (S/Irr) types of satellites
around bright host galaxies.
We made a volume-limited sample of 4,986 satellites brighter than 
$M_r = -18.0$ associated with 2,254 hosts brighter than $M_r =-19.0$
from the SDSS DR5 sample. The morphology of satellites is determined by
an automated morphology classifier,
but the host galaxies are visually classified. 
We found segregation of satellite morphology as a function of
the projected distance from the host galaxy.
The amplitude and shape of the early-type satellite fraction profile
are found to depend on the host luminosity.
This is the morphology-radius/density relation at the galactic scale.
There is a strong tendency for morphology conformity 
between the host galaxy and its satellites.
The early-type fraction of satellites hosted by early-type galaxies is
systematically larger than that of late-type hosts, and is
a strong function of the distance from the host galaxies.
Fainter satellites are more vulnerable to the morphology transformation
effects of hosts.
Dependence of satellite morphology on
the large-scale background density was detected. 
The fraction of early-type satellites increases in high
density regions for both early and late-type hosts.
It is argued that the conformity in morphology of galactic satellite system
is mainly originated by
the hydrodynamical and radiative effects of hosts on satellites.
\end{abstract}

\section{Introduction}

Morphology reflects the integral property of a galaxy, such as stellar
populations, gas content, and dynamical structures. Its origin is one of
the central problems in the study of galaxy formation and evolution. 
If a galaxy remains isolated after its formation, 
all of its physical properties would be entirely
determined by the initial conditions of the proto-galactic cloud
and by the subsequent internal evolution.
But, it seems unlikely 
because galaxies are believed to form through a series of minor/major mergers.
In fact, the isolated bright galaxies in high density regions 
are more likely to be recently merged ones and the
morphology of galaxies contains imprints of interaction with environment 
in addition to the formation process (Park, Gott, \& Choi 2008). 

There is observational evidence that shows an intimate correlation
between the morphology of the central galaxy and its neighbors (Wirth 1983;
Hickson et al. 1984; Ramella et al. 1987; Osmond \& Ponman 2004;
Weinmann et al. 2006; Park et al. 2008). 
Recent analysis of the morphology of Sloan Digital Sky Survey (SDSS;
York et al. 2000) galaxies 
by Park et al. (2007, 2008) showed that
galaxy morphology does depend on the large-scale background density
but the role of the nearest neighbor is more decisive.
The critical roles of the closest neighbor in determining galaxy morphology
appear as the galactic conformity (Weinmann et al. 2006) 
between a galaxy and its neighbors.
Galaxy morphology also depends on luminosity in 
that galaxy morphology is more likely to be early type for brighter galaxies. 
Since bright galaxies mainly live in high density regions
through the luminosity-density relation, it appears that early types are 
more prevalent at high densities.

Satellite systems are good places to inspect the environmental dependence
of galaxy morphology and to study the galaxy formation process since 
they are abundant and very localized systems with a size of less 
than 1 Mpc.
Most of the previous studies of satellite galaxies
were focused on the radial distribution of satellite galaxies (Sales \& Lambas 
2005; van den Bosch et al. 2005; Chen et al. 2006), 
dark matter halo (McKay et al. 2002; Prada et al. 2003; van den Bosch 2004), 
and angular distributions (Zaritsky et al. 1997; Sales \& Lambas 2004;
Zentner et al. 2005; Yang et al. 2006; 
Libeskind et al. 2007; Bailin et al. 2007;
Kang et al. 2007; Sales et al. 2007).
The morphology of satellite galaxies is also an
observable parameter that is directly related to 
formation and evolution of galaxy.

The purpose of the present paper is to study the relation between 
the morphology of satellite galaxies and the local environment
such as the host morphology and background density. 
We used large and homogeneous morphology samples made by both 
visual and automated classifications.
We will see a tight correlation between the 
host and satellites morphologies.
The satellite systems in our study are hosted by the typical bright galaxies, 
and are not in general large groups or clusters of galaxies. 
Our host sample is dominated
by the $L_*$ galaxies, and their satellites are fainter 
by about two magnitudes.

\section{Data}
\subsection{Isolated satellite systems}

The basic source of data is the large-scale structure sample (LSS), DR4plus, 
from the New York University Value-Added Galaxy Catalogue (NYU-VAGC;
Blanton et al. 2005) which is a subset of the SDSS Data 
Release 5 (DR5; Adelman-McCarthy et al. 2007).
The primary sample of galaxies used here is a subset of the LSS-DR4plus, 
which includes Main galaxies (Strauss et al. 2002) with extinction 
corrected apparent Petrosian $r$-magnitudes in the range 
$14.5 \leq r_{\rm Pet} < 17.77$ and redshifts in the range 
$0.001 < z < 0.5$. 
Our survey region covers 4464 deg$^2$, which 
is shown in Figure 1 of Park et al. (2007). To this primary sample, 
we added the galaxies brighter than the bright limit 
($r_{\rm Pet}=14.5$) of the sample. 
Various existing redshift catalogs are searched for the redshifts of
the bright galaxies with no spectrum.
The catalogs include RC3 (de Vaucouleurs et al. 1991), 
Catalog of Nearby Galaxies (Tully \& Fisher 1988) and 
Updated Zwicky catalog
\footnote{http://www.cfa.harvard.edu/\~{}huchra/zcat/zcom.htm}
(ZCAT 2000 Version). In case of no 
measured redshift even in these catalogs, we used the redshift taken 
from NASA/IPAC Extragalactic Database\footnote{
http://nedwww.ipac.caltech.edu/} (NED) when available.
We added 5,503 bright galaxies to the primary sample.
The final data set consists of 370,789 galaxies with known redshift
and photometry.
Throughout this paper, we use a flat $\Lambda$CDM cosmology with density
parameters $\Omega _m = 0.27$ and $\Omega_{\Lambda} = 0.73$.

To search for isolated satellite systems we take two steps.
We first look for isolated galaxies in a volume-limited sample of
galaxies brighter than the $r$-band absolute magnitude $M_r = -19.0
+ 5 {\rm log} h$ (hereafter we are going to drop the term $5 {\rm log} h$) and
with redshifts between 0.02 and 0.04724.
The lower redshift limit
is chosen to make our sample as complete as possible since galaxies  with
$z < 0.02$ in the SDSS seems to be incomplete even though
we supplemented bright galaxies (Park et al. 2007).
The comoving space number density of galaxies is approximately constant in
the radial direction at $z>0.02$, but drops significantly at $z<0.02$.
The upper limit of $z=0.04724$ corresponds to the survey limits of $r=17.77$ 
for a galaxy with $M_r = -18.0$.

A target galaxy is isolated if the projected separation to its nearest
neighbor galaxy is larger than the virial radii of both galaxies.
The neighbors of a target galaxy with $M_r$ are those with absolute 
magnitude brighter than $M_r +1.0$, and velocity difference less than
1,000 km s$^{-1}$. 
We have also used the most influential neighbor instead of the nearest one
for a comparison in the measurement of  the projected separation $r_p$. 
Our results are basically the same for these two choices.
The most influential neighbor is the neighbor which
induces the highest local density at the location of the target galaxy.
Given $r_p$ between them, we calculate the local
mass density due to the neighbor with luminosity $L_n$ by
\begin{equation}
\rho_{n}= 3\gamma_n L_n /4\pi r_p^3,
\end{equation}
where we adopt the mass-to-light ratios $\gamma_n=2$ for early types and
1 for late types.
This choice is based on the morphology-specific central stellar velocity 
dispersion and on the pairwise peculiar velocity difference of 
early and late-type galaxies with their neighbors 
(see Park et al. 2008 for more details).
We define the virial radius of each galaxy as the radius where the mean 
density within the sphere centered at the galaxy given by equation (1) 
becomes the virialized 
density, which is set to $766{\bar\rho}$ (see section 3.1 of Park et al.
2008). The mean mass density is obtained from
${\bar{\rho}}= \sum_{all} \gamma_i L_i/V $ where the summation is over
all galaxies in our full volume-limited sample of volume $V$
with the absolute magnitude constraint $M_r <-18.0$.
An early or late-type galaxy with $M_r=-20$ has virial radius of
300 or 240 $h^{-1}$kpc, respectively.
For those with $M_r = -20.5$, the virial radii are 350 and 280 
$h^{-1}$kpc, respectively.

We found 8,883 isolated galaxies in our volume-limited sample. They are
physically isolated ones in the sense that they are not hydrodynamically
interacting with neighbors. In all previous studies isolation is
determined by using a pre-selected fixed radius ignoring the physical
size of individual galaxies involved. A blindly large radius of 
the isolation boundary results in too small sample size, while any
fixed value in the right range results in contamination in the sample
with interacting galaxies added.

Once the bright isolated galaxies are found, we search for satellites
associated with them. We limit the satellite candidates only to 
galaxies with $M_r$ brighter than $-18.0$, a limit one magnitude fainter 
than that of the host candidates. This choice gives us a uniform and 
complete selection of satellites for host galaxies also uniformly and 
completely selected across our sample volume (see Figure 1 below). 
At each location of the isolated galaxies
we search for galaxies with velocity difference less than 500 km s$^{-1}$,
absolute magnitude more than one magnitude fainter (but brighter than
$-18.0$), and the projected separation less than the smaller of
1 $h^{-1}$Mpc and $r_p$(neighbor)$- r_{vir}$(neighbor), where the letter is
the difference between the host-neighbor separation and the neighbor's 
virial radius.
We used the Petrosian $g$-band absolute magnitude for satellite identification
because $g$ magnitude is 
most similar to the $B$ magnitude that is used for some bright galaxies whose
SDSS photometry is too poor to be used without correction.

\begin{figure}
\includegraphics[scale=0.4, bb=20 0 570 650, clip]{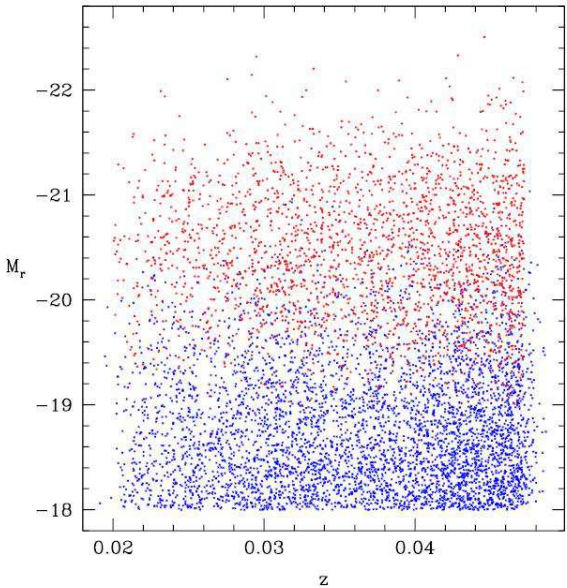}
\caption{ Our volume-limited sample of the host galaxies (red points) 
and satellites (blue).  The host distribution shows our sample
boundaries in redshift. Some of the satellites lie beyond 
the redshift boundaries of the volume-limited sample 
because we allowed 500 km$s^{-1}$ difference in radial velocity 
between a host and its satellites.}
\end{figure}

Among the 8,883 isolated galaxies, 2,254 have satellites, and the total
number of satellites belonging to these systems is 4,986. 
Figure 1 shows the distributions of the host galaxies (red points) and
satellites (blue).  The host distribution shows our sample
boundaries in redshift. Some of the satellites lie beyond 
the redshift boundaries because we allowed 500 km$s^{-1}$ difference
in radial velocity in the search for satellites. The median absolute 
magnitude of the hosts is 
$M_r = -20.47$, which is very close to that of the $L_*$ SDSS galaxies
(Choi et al. 2007). Therefore, the host galaxies of our satellite systems
are dominated by normal bright galaxies, and are not in general
the central $cD$ galaxies holding the bright galaxies as satellites.
The median absolute magnitude  of our satellites is $-18.67$.
So they are not dwarf galaxies, but subluminous bright galaxies 
typically 1.8 magnitude fainter than their hosts.
Since both our hosts and satellites are selected uniformly in the absolute
magnitude space, our study of satellite morphology is unbiased against
host and satellite luminosity.

\subsection{Morphology}

We classify the morphology of host galaxies by the visual inspection
because visual classification is accurate for bright galaxies. 
For visual classification, ellipticals(E) and lenticulars(S0) as well as 
spiral(S) and irregulars(Irr) are distinguished, 
but for better statistics we categorized 
E and S0 galaxies as early types, and S and Irr galaxies 
as late types in the present analysis.
We mainly employed the automated classifier of Park \& Choi (2005)
for satellites. This classifier divides galaxies into 
early and late types based on their location in the three-dimensional
parameter space of $u-r$ color, $g-i$ color gradient, and 
the $i$-band concentration index. The classification boundaries 
in the parameter space are
chosen by using a large training set of galaxies with known morphology.
All of the satellites are visually checked.
But the visual classification is used only as a complementary one,
especially for relatively bright satellites or 
for those undergoing close interactions or mergers. 
This is because in most cases ($>$ 90\%) the visual and automated
classifications of satellites agree with each other, and because
for the very faint galaxies 
close to the faint limit of the sample, it is not certain whether or not
the visual classification is on average any better than the result of the
automated classification.

\section{PROPERTIES OF SATELLITE SYSTEMS}

\subsection{Morphology and radial distribution of satellites}
 
We measured the early-type fraction $f(E_{s})$ and surface number density 
$\Sigma(E{_s})$ of satellite galaxies 
as a function of projected distance ($r_{p}$) from the host galaxies. 
The top panel of Figure 2 shows the early-type fraction of satellites
associated with our isolated early-type hosts $f(E_{s}|E_{h})$ (filled circles) and
isolated late-type hosts $f(E_{s}|L_{h})$.
The innermost bin is $r_{p} < 37.8 h^{-1}$ kpc, which corresponds
to the fiber collision radius of $55^{\prime \prime}$ at the outer boundary
($z=0.04724$) of our volume-limited sample.
\begin{figure}
\includegraphics[scale=0.55, bb=20 135 570 750, clip]{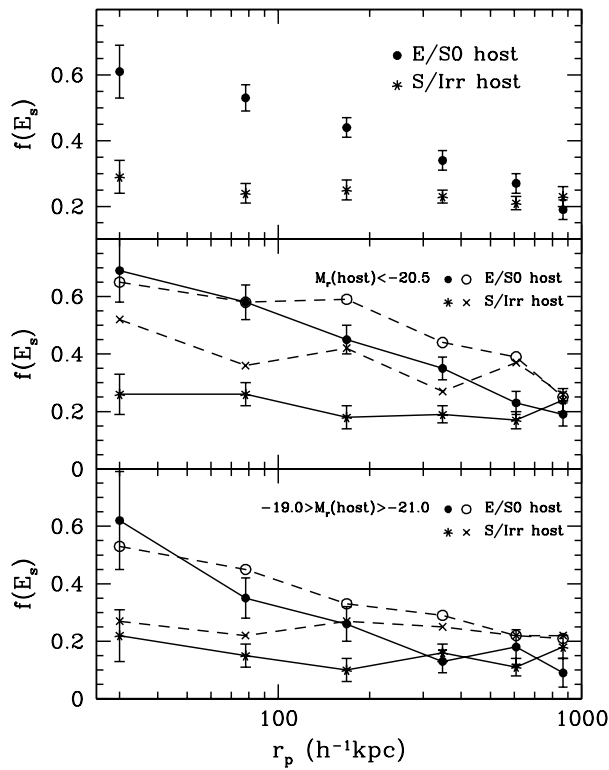}
\caption{Early-type fraction of satellite
galaxies as a function of projected distance from host galaxies.
In the top panel satellites are divided into those associated with
early-type (filled circles) and late-type (stars) hosts. In the middle
and bottom panels satellites are further divided into those having
magnitude difference with hosts greater than 1.9 (solid lines) and
less than 1.9 but more than 1.0 (dashed lines).
}
\end{figure}
It can be noted that $f(E_{s}|E_{h})$ is significantly higher
than  $f(E_{s}|L_{h})$ at least out to about 350 $h^{-1}$kpc,
which is roughly the virial radius of the typical early-type host galaxies
analyzed in this study.
This result means that the morphology of satellites tends to be similar
to that of hosts. It demonstrates 
the morphology-radius relation at the galactic scales.
A similar finding was reported by Weinmann et al. (2006) for galaxies
in groups and clusters, and by Park et al. (2007, 2008) for galaxy pairs.
For late-type hosts, the early-type satellite fraction increases
very slowly as satellites approach their hosts. 
Some of this effect must be due to the morphology-luminosity
relation. The early-type hosts are in general brighter than the late-type
hosts, and correspondingly the satellites of our early-type hosts 
are also on average brighter than those of late-type hosts
due to our satellite finding process, i.e. more than one magnitude
fainter relative to the host.
Because of the morphology-luminosity relation, the morphology of
early-type hosts' satellites is in general earlier than that of
late-type hosts' satellites even if there is no direct physical 
influence of the host on satellites.
We do not think this is the main reason for the host-satellite
morphology correlation we found because the satellite morphology is a very
strong function of host-satellite separation in early-type host systems
and the early-type satellite fractions for early and late-type hosts
start to merge at $r_p \sim 1h^{-1}$Mpc.

Irrelevance of the morphology-luminosity relation to our findings can be also
demonstrated by the early-type satellite fraction plot 
drawn for hosts with fixed luminosity.
The middle and bottom panels of Figure 2 show the early-type satellite fractions 
for host galaxies brighter than $M_r=-20.5$ and fainter than $-21.0$,
respectively.  We allowed an overlap in $M_r$ to decrease the statistical
fluctuations.
We also divided satellites into
a subset more than $\Delta M_g=1.9$ magnitude fainter than the host and
a subset more than 1.0 but less than 1.9 magnitude fainter.
Drawn are the four cases of early-type hosts ($E_h$) and 
satellites with $\Delta M_g >1.9$
(filled circles, solid line), $E_h$ and satellites with $1.0<\Delta M_g<1.9$ 
(open circles, dashed line), $L_h$ and satellites with $\Delta M_g >1.9$
(stars, solid line), and $L_h$ and satellites with $1.0<\Delta M_g<1.9$
(crosses, dashed line).

The satellites with smaller $\Delta M_g$ are on average brighter than those
with larger $\Delta M_g$, and are more likely to be early-types in accordance
with the morphology-luminosity relation. The mean level of $f(E_s )$ at
very large $r_p$ is indeed higher for smaller $\Delta M_g$ satellites in
both middle and bottom panels of Figure 2.
Once we subtract the dependence of this asymptotic value on host and satellite
luminosity from these figures, interesting dependence of $f(E_s )$ on $r_p$ and host
morphology becomes evident.
The fraction of early-type satellites associated with early-type hosts,
$f(E_s | E_h)$, depends on $r_p$ more sensitively for fainter satellites 
(compare the open and filled circles). This is true for both relatively 
bright (middle panel) and faint (bottom panel) hosts.
It can be also noted from the middle and bottom panels that
the outer boundary of the region of early-type host influence is farther
for brighter hosts. The net effects of the late-type hosts on satellite
morphology seem insignificant.

The satellites of early-type hosts are likely to be deprived of their 
cold gas through the hydrodynamic and radiative interactions with 
the X-ray emitting hot gas of their host. The satellites of 
late-type hosts are in principle able to get cold gas from their hosts 
although the hot gas in the halo of late-type hosts can also remove 
the cold gas in their satellites. 
Based on a detailed study of morphology-environment relation of galaxy pairs
Park et al. (2008) concluded that the galaxy morphology-local density
relation is mainly due to the interaction between nearest neighbor
galaxies. When galaxies are closer than their virial radii, they start
to interact hydrodynamically and this causes the conformity in morphology
of close galaxy pairs.  
The present results support their scenario, and this seems to be the origin
of the morphology conformity in galactic satellite systems.

One major difference between our result and that of Park et al. is that
the satellite morphology does not tend to be of late type as satellites
approach late-type host. The galaxy pairs in Park et al.'s sample 
are dominated by those with similar luminosity and therefore their 
interaction can affect physical properties of both galaxies significantly. 
On the other hand, in the current analysis 
satellites are typically 1.8 magnitudes fainter than hosts,
and the influence is largely lopsided from hosts to satellites.
The slight rising tendency of $f(E_{s})$ 
very close to late-type hosts can be because satellites are suffering from cold
gas stripping and ionization by the host halo gas, but can not actively 
catch the cold gas from their hosts as efficiently as companion galaxies
having luminosity similar to the hosts.

The slopes of the surface density profiles also reflects the physical
effects of their host galaxies on satellites.
Figure 3 presents the satellite surface number
density profiles for early (top panel) and late-type hosts.
The ratio of two profiles in each panel gives the morphology
fraction in the top panel of Figure 2.
\begin{figure}
\includegraphics[scale=0.4]{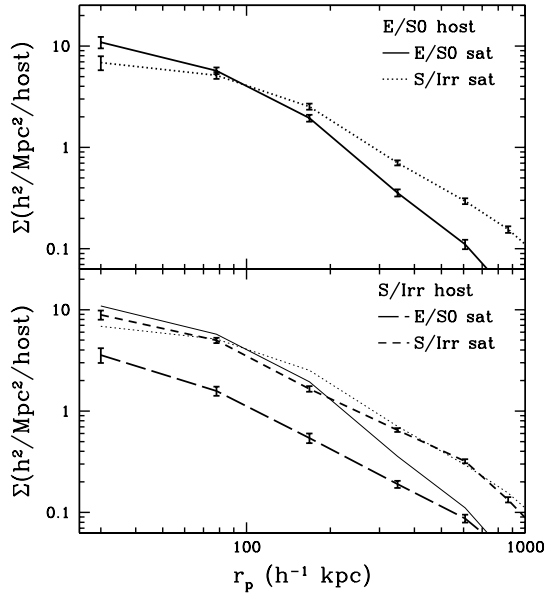}
\caption{Surface density distribution of satellite 
galaxies as a function of projected distance from the host galaxies
with $M_r < -19.0$.
The surface density profiles of satellites hosted
by the early-type galaxies are repeated as thin lines in the bottom panel.
}
\end{figure}
It demonstrates the number density profile of satellites critically
depends on both host and satellite morphology.
The surface density profiles
show why $f(E_{s}|E_{h})$ decreases more rapidly than $f(E_{s}|L_{h})$. 
It is due to the dramatic drop in the surface density
of early-type satellites hosted by early-type hosts $\Sigma(E_{s}|E_{h})$,
and to the slower drop of that of late-type satellites hosted 
by early-type hosts $\Sigma(L_{s}|E_{h})$. 
This makes late-type satellites dominant in the satellite systems of 
early-type hosts at $r_{p} > 100\sim 200$ $h^{-1}$ kpc,
where the exact location depending on the host luminosity and
the host-satellite magnitude difference (see Figure 2).
In the systems hosted by late-type galaxies, late-type satellites
are dominant at all $r_p$. The shapes of the surface density profiles 
of both early and late-type satellites, 
$\Sigma(E_{s}|L_{h})$ and $\Sigma(L_{s}|L_{h})$,
are similar to each other, making their ratio roughly constant
of $r_p$. At large separations, $r_p > 600 h^{-1}$kpc,
the surface densities satellites belonging to
early-type (upper panel) and late-type (lower panel) hosts approach 
roughly the same ratio, resulting in $f(E_s)\approx 0.2$.
This seems the field value of galaxy morphology for galaxies
having absolute magnitudes similar to those of the satellites in our sample.

The morphology fraction shown in Figure 2 is the result of
projection of the three dimensional distribution on the sky.
In order to get a rough idea on the central morphology fraction
we try to deproject the profile as follows.
We assume the radial number density of each of early and late-type satellites
follows a power-low, $\rho(r)=\rho_o (r/r_{o})^{-\gamma}$.
Then the projected density follows the form (Binney \& Tremaine 1987),
\begin{equation}
\Sigma(r_{p})=\rho_{o}r_{o}^{\gamma} (-{1\over2})!({{\gamma-3}\over{2}})!/
r_{p}^{\gamma-1}({{\gamma-2}\over{2}})!.
\end{equation}
The parameters in the fraction are obtained from a least-square fit to the 
inner-most three points shown in Figure 3 for each case of host
and satellite morphology.
Only two parameters are free. We found the slope of the three-dimensional
profile is $-1.8\sim-1.9$ at $r<200 h^{-1} kpc$ 
except for the late-type satellite associated with
early-type host case, which has about -1.5. The true fraction of early-type
satellites very close to early-type hosts is found to reach about
0.71 and 0.78 at $r=30$ and 10$h^{-1}$kpcs, respectively.
On the other hand, the fraction for late-type hosts in 3D is nearly
the same as that shown in Figure 2 because the slope of radial density
profile is almost independent of satellite morphology in this case.

\subsection{Background density dependence}

As argued in the previous sections, the morphology conformity in galactic 
satellite systems seems to be due to the local effects
of hosts on their satellites. However, the galactic 
conformity can be affected by the global environment as well as local one.
Park et al. (2008) showed that, even though the morphology of galaxies
depends mainly on luminosity and the small-scale environment due to the nearest
neighbor, it also depends on the large-scale background density.
The dependence of galaxy morphology on the large-scale density was found
even when both the luminosity of the target galaxy and the environment due to
the nearest neighbor were fixed.
This was explained by the dependence of the
hot halo gas of galaxies on the large-scale density.
In this section we look for a similar effect on galactic satellites. 

We used the galaxy number density estimator defined by
20 nearest $L_*$ galaxies with $-20.0 > M_r > -21.0$
drawn from the full volume-limited sample
\begin{equation}
\rho_{20}/{\bar{\rho}}= \sum_{i=1}^{20} W_i(|{\bf x}_i -
{\bf x}|)/{\bar\rho},
\end{equation}
where  $W(r)$ is a spline-kernel weight 
and $\bar{\rho}$ is the mean number density of the $L_*$ galaxies
in the SDSS. This choice is the same as those used by Park et al. (2007).
The median value of the effective Gaussian 
smoothing scale, corresponding to the adaptive spline smoothing, 
is $4.7 h^{-1}$ Mpc.

The top panel of Figure 4 shows the distributions of the large-scale density for 
early-type (solid line) and late-type (dotted) hosts. It can be seen that their
distributions are nearly the same except for the highest densiy bin 
even though the early-type galaxies are in general
preferentially located at higher densities. This may be because the isolation
constraint on hosts excluded more early-type galaxies than late-types 
in high density regions.
However, as can be seen in the bottom panel of Figure 4, our isolated host
sample still respects the luminosity-density relation.

\begin{figure}
\includegraphics[scale=0.55, bb=10 135 570 600, clip]{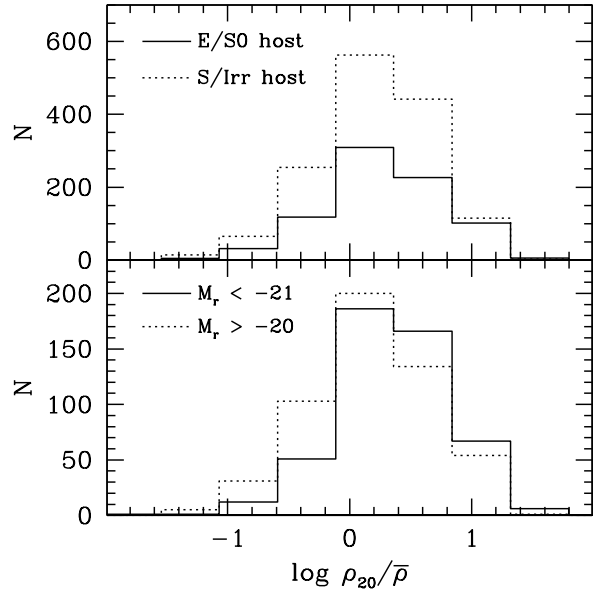}
\caption{Distributions of the large-scale background density
at the location of our isolated host galaxies, divided into morphology 
subsets (upper panel) and luminosity subsets (lower panel).
}
\end{figure}

\begin{figure}
\includegraphics[scale=0.55, bb=10 145 570 700, clip]{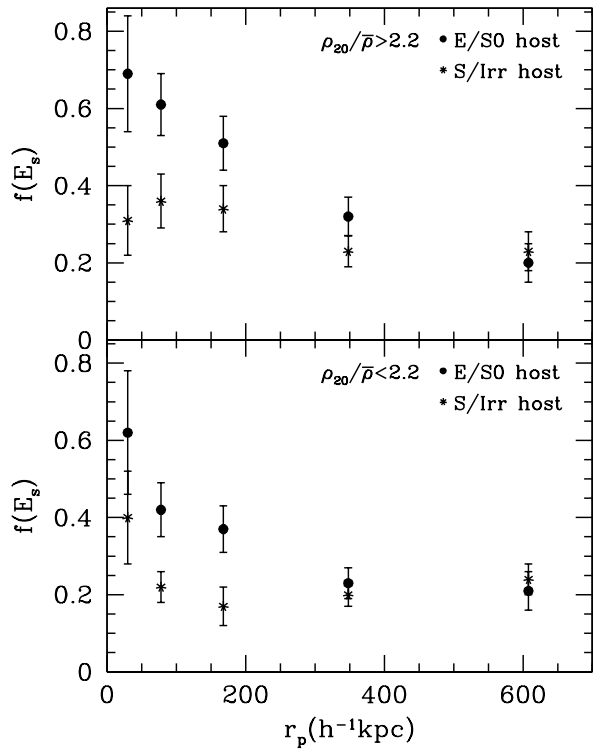}
\caption{Early-type fractions of satellite galaxies as a function of
the projected distance from host galaxies. Top panel is for high density regions
and the bottom panel for low density regions, respectively.
The luminosity of hosts is fixed to $-20.5>M_{r}>-21.5$.}
\end{figure}

Figure 5 shows $f(E_{s})$ as 
a function of the projected separation from the hosts in different 
large-scale background density regions.
We fixed the luminosity of host galaxies to
$-20.5>M_{r}>-21.5$ to separate the luminosity effects from the background
density effects on galaxy morphology.
The large scale environment is divided into high and low 
density regions with $\rho_{20}/{\bar{\rho}} > 2.2$ and $< 2.2$, 
respectively, where  $\rho_{20}/{\bar{\rho}} = 2.2$ is the median density
for our isolated hosts.

It can be seen in Figure 5 that $f(E_{s})$ is higher in high density
regions for both early and late-type hosts with fixed luminosity.
The background density seems to play a definite role in determining the 
morphology of galactic satellites.
The background density can directly affect 
the satellites, or affect them indirectly through the host whose properties
depend statistically on the background density.
Park et al. (2008) found that 
the early-type galaxies in high density regions have higher X-ray 
luminosity than those in low density regions even when their optical 
luminosity is the same.
This means that the hot halo gas of early-type galaxies is hotter and
denser at high densities.
Taking into account this finding 
we interpret the background density dependence of the satellite morphology
as due to the hydrodynamic and radiative effects of the hot gas of 
host galaxies on satellites.
This is supported by the fact that, even though $f(E_s)$ is generally
higher in high density regions in Figure 5, it is so only when satellites 
are close to their hosts and $f(E_s)$ at $r_p$ much larger than the host's
virial radius is rather independent of the background density.
If the background density directly affect the morphology
of satellites, the satellite morphology should depend on the background density
at all host-satellite separations.
It can be also noted that $f(E_{s})$ is higher for early-type hosts
than for late-type hosts both in high and low density regions. 
Therefore, the conformity in morphology at galactic scales prevails
in both high and low density environments.

The $f(E_{s})$ in high density environment decreases almost linearly
with $r_{p}$ while $f(E_{s})$ in low density environment decreases nearly
exponentially. It is surprising to see this background density dependence
even if we fixed host morphology and luminosity. All $f(E_{s})$ seems to converge at 
$r_{p}>600$ $h^{-1}$ kpc. This is a scale a little larger than the virial 
radius of the host galaxies under consideration.
The virial radii are about 280 and 350 $h^{-1}$kpc
for late and early-type galaxies with $M_r = -20.5$.
In the analysis of luminous galaxy pairs the dependence of galaxy morphology
on the neighbor's morphology appears at separations of $r_p \la r_{\rm vir}$
(Park et al. 2008).
This was explained by the hydrodynamic interactions between the pairs
within the virialized region.


Previous studies showed that the fraction of interlopers could be large
at large $r_p$ (Prada et al. 2003) and that the interloper fraction
depends on the color of the satellites, with interlopers being
rare amongst the red satellites, but making up about half of the blue satellites.
If our satellite samples were dominated by interlopers at large $r_p$, 
the difference in $f(E_s)$ in high and low density regions 
could be due simply to the interlopers which
respect the morphology-density relation. However, Figure 5 shows that
$f(E)$ converges to about 0.2 both in high and low density regions
and both for early- and late-type host galaxies.
Therefore, the satellites at large separations do not show the trends
that are expected for the general background galaxies.
This indicates our results are not significantly affected by interlopers.

\section{Discussion and Conclusions}

We have found the morphology-radius relation for galactic satellite
systems. Early-type satellites are prevalent in the vicinity of early
type hosts. 
The origin of the conformity in morphology is thought to be
the hydrodynamic and radiative effects of hosts on satellites
in addition to the tidal (gravitational) effects.

The satellite morphology is found to depend on the large-scale background density. 
In high density regions the early-type fraction of satellites decreases
relatively slowly beyond the virial radius of the host galaxy.
However, in low density regions
the fraction of satellites with early morphological type 
drops sharply at separations of $r_p =50 \sim 200 h^{-1}$kpc 
for both early and late-type host systems.
As we fixed the mass of host galaxies by fixing luminosity and morphology,
this difference must be coming from non-gravitational effects.
It is argued that the hot halo gas of the host galaxies is responsible 
for prevalence of early-type satellites in the vicinity of hosts, and that
in high density regions the hot halo gas
can be more confined by the ambient intergalactic medium and
has higher density and temperature, which can better deplete the cold gas
in satellites more efficiently.

The galactic conformity found from the present sample of satellite systems
is not much affected by the detailed selection criteria of the satellites. 
The magnitude difference between host and satellites is not critical
because we obtained similar results for the satellite systems defined by 
different magnitude differences. 
Using the most-influential neighbors instead of the nearest neighbors
in identification of isolated hosts and satellites also did not make
much difference.
We also examined whether or not our
results are affected by our isolation requirement for host galaxies,
and found that all of our results qualitatively remain the same.
We made exactly the same analysis for satellites defined for host galaxies
which are not constrained to be isolated. In this analysis
a galaxy becomes a satellite if it finds a host galaxy within $r_p=800 h^{-1}$kpc
that is more than 2 magnitudes brighter and has velocity difference less 
than 500 km s$^{-1}$. If there is more than one such hosts, the closest
one is chosen. Hosts are limited to $M_r <-20.0$, and satellites have
$M_r < -18.0$. We found 8,353 satellites in 3,472 systems. We obtain 
basically the same results for these satellite systems as for the isolated
ones but with much higher statistical significances.
Therefore, our results are robust against various choices of parameters used
to identify hosts and satellites.

In the forthcoming paper we will study the shape and internal properties of satellites. 
Rather than dividing satellites into early and late types, we will adopt a new
classification scheme that is more appropriate for the satellite galaxies.
We found this is necessary because our satellite galaxies are fainter and 
located in the special environment given by the hosts compared to the normal
bright galaxies for which the usual morphology classification schemes are
developed.

\section*{Acknowledgments}
H.B.A., Y.Y.C. and C.B.P. 
acknowledge the support of the Korea Science and Engineering
Foundation (KOSEF) through the Astrophysical Research Center for the
Structure and Evolution of the Cosmos (ARCSEC).

Funding for the SDSS and SDSS-II has been provided by the Alfred P. Sloan Foundation, the Participating Institutions, the National Science Foundation, the U.S. Department of Energy, the National Aeronautics and Space Administration, the Japanese Monbukagakusho, the Max Planck Society, and the Higher Education Funding Council for England. The SDSS Web Site is http://www.sdss.org/.

The SDSS is managed by the Astrophysical Research Consortium for the Participating Institutions. The Participating Institutions are the American Museum of Natural History, Astrophysical Institute Potsdam, University of Basel, University of Cambridge, Case Western Reserve University, University of Chicago, Drexel University, Fermilab, the Institute for Advanced Study, the Japan Participation Group, Johns Hopkins University, the Joint Institute for Nuclear Astrophysics, the Kavli Institute for Particle Astrophysics and Cosmology, the Korean Scientist Group, the Chinese Academy of Sciences (LAMOST), Los Alamos National Laboratory, the Max-Planck-Institute for Astronomy (MPIA), the Max-Planck-Institute for Astrophysics (MPA), New Mexico State University, Ohio State University, University of Pittsburgh, University of Portsmouth, Princeton University, the United States Naval Observatory, and the University of Washington. 

{}

\label{lastpage}
\end{document}